
\magnification1200
\baselineskip=12truept

\newcount\fcount\fcount=0
\def\ref#1{\global\advance\fcount by 1 \global\xdef#1{\relax\the\fcount}}
\vsize=22 truecm
\hsize=16.5 truecm
\parskip=0.2cm
\parindent=0truecm
\raggedbottom
\def\c{\centerline}

\def\bs{\bigskip}


\tolerance=10000

\def\today{\number\year\space \ifcase\month\or 	January\or February\or
	March\or April\or May\or June\or July\or August\or September\or
	October\or November\or December\fi\space \number\day}


\def\pp{\par\hangindent=.125truein \hangafter=1}
\def\aref#1;#2;#3;#4;#5;#6.{\item{#1.} #2 (#3). {\it #4} {\bf #5}, #6.}
\def\abook#1;#2;#3;#4;#5;#6;#7.{\item{#1.} #2 (#3). in {\it #4}, #5, #6, #7.}
\def\apress#1;#2;#3;#4;#5.{\item{#1.} #2 (#3). {\it #4}, #5.}
\def\arep#1;#2;#3;#4;#5.{\item{#1.} #2 (#3). #5, #4.}
\newcount\fcount \fcount=0
\def\ref#1{\global\advance\fcount by 1
  \global\xdef#1{\relax\the\fcount}}
\def\spose#1{\hbox to 0pt{#1\hss}}
\def\simlt{\mathrel{\spose{\lower 3pt\hbox{$\mathchar"218$}}
     \raise 2.0pt\hbox{$\mathchar"13C$}}}
\def\simgt{\mathrel{\spose{\lower 3pt\hbox{$\mathchar"218$}}
     \raise 2.0pt\hbox{$\mathchar"13E$}}}
\def\etal{{\it et al.}}

\def\frac#1/#2{\leavevmode\kern.1em
 \raise.5ex\hbox{\the\scriptfont0 #1}\kern-.1em
 /\kern-.15em\lower.25ex\hbox{\the\scriptfont0 #2}}

\parindent=.5truecm
\parskip=3pt

\line{\hfil CfPA-TH-95-15}
\line{\hfil astro-ph/9508040}
\line{\hfil \today}
\bigskip

\c{\bf  THE FIRST GENERATION OF STARS:}
\c{\bf FIRST STEPS TOWARDS CHEMICAL EVOLUTION OF GALAXIES}
\bs\bigskip

\c{ Jean Audouze }
\smallskip
\centerline{\it Institut d'Astrophysique du CNRS}
\centerline{\it 98 bis Boulevard Arago, 75014 Paris, France}
\c{and Joseph Silk}
\smallskip
\centerline{\it Center for Particle Astrophysics}
\centerline{\it and Department of Astronomy,}
\centerline{\it University of California, Berkeley, CA 94720-7304}

\bigskip
\bigskip

{\bf Abstract}
\smallskip
{\noindent
  We argue that extreme metal-poor stars show a high dispersion in metallicity,
 because  their abundances are the outcome of very few supernova events.
Abundance anomalies should appear because of the discrete range of progenitor
masses. There is a natural metallicity threshold of $Z/Z_\odot\sim 10^{-4}$
below which
one would  expect to find very few, if any, halo stars. Similar reasoning is
applied to
lower mass systems, such as metal-poor compact blue galaxies and Lyman alpha
absorption line clouds seen towards high redshift quasars, where a somewhat
higher threshold is inferred.}

{\it Subject headings: } Galaxy: formation---Galaxy: halo---stars:abundances
\vfill

\eject

{\bf 1. Introduction}

A recent analysis  (McWilliam, Preston, Sneden and Searle 1995) of extremely
metal-poor stars to $\rm [Fe/H]\simgt -4$, found
remarkable trends concerning many abundances, including those of alpha-elements
(Mg, Si, Ca), iron peak
elements (Sc, Ti, Cr, Mn, Co, Ni) and heavy elements (Y, Sr, Ba, Eu),
with respect to [Fe/H],
that appear over the range $\rm [Fe/H]=-4$ to $-2.4.$  A large abundance
dispersion was also found
between different stars, for the most metal-poor stars ($\rm [Fe/H]\simgt -4$).
 We shall argue that these data provide novel clues about the nucleosynthetic
role of the first generation of stars. We interpret these variations and trends
as the signature of an epoch of  galactic evolution before mixing of stellar
ejecta had time to be efficient.  This constraint sets a metallicity threshold,
we shall show, of $\rm [Fe/H]\approx -4.$ We also discuss similar abundance
trends observed in blue compact galaxies and high redshift metal-poor quasar
absorption line systems.
\bs
\goodbreak

{\bf 2. The first generation of galactic stars}

A general timescale argument for the early enrichment is as follows: the star
formation rate in the disk of our galaxy is constant, to within a factor of 2,
over the past 10 Gyr. In a simple closed box model, one would expect, over the
initial lifetime of a single generation of massive stars of approximately
$2\times 10^6$ yr,  to attain a metallicity of roughly $2\times 10^{-4}$ of the
solar value. This implies that only one generation of
massive stars had time to explode as Type II supernovae and eject
metal-enriched material into the interstellar medium from which these extremely
metal-poor stars were formed.

An estimate of the first metal enrichment is as follows.
The number of Type II  supernovae that would have exploded in this initial
period should be $N \sim 1-2 \times 10^4$,
if their rate is comparable to the  present rate of Type II supernovae, namely
 1 per 100 yr. This should be a conservative estimate, since considerations
of the diminished role of mass loss for zero-metal stars (Jura 1986)
and of prompt initial enrichment (Truran and Cameron 1971)
suggest that the rate of Type II supernovae may have been higher in the early
galaxy.
Assuming that each Type II  supernova ejects approximately
 $1 M_\odot$
of metals, we infer a mean  initial enrichment of $Z=3\times 10^{-6} M_{10},$
where the mass in  Population II is written as $10^{10}M_{10} M_\odot.$

However we note that the interstellar medium enrichment will be highly
inhomogeneous during this early phase of the galaxy. The reason for this is
that the mean separation between supernova remnants is
$(V/N)^{1/3},$ where $V$ is the volume of the spheroid, $\sim 10^{11}\rm pc^3,$
or about 100 to $200\,\rm pc.$ The maximum extent of a supernova remnant before
breakup and dispersion into the interstellar medium should be about half the
mean separation. We may utilize theoretical calculations of the evolution of
radiative supernova remnants evolving in a homogeneous interstellar medium
(Cioffi, McKee and Bertschinger 1988) to estimate a mean size
when the remnant finally merges with the ambient medium of
$69[E_{51}^{0.32}n^{-0.37}\sigma_{10}^{-0.43}\zeta^{0.09}]\rm pc.$ Here
$E_{51}$ is the initial kinetic energy of the supernova remnant in units of
$10^{51}$ ergs, $n$ is the ambient density of the interstellar medium in
particles cm$^{-3}$, $\sigma_{10}$ is the velocity dispersion in units of
$10\,\rm km\,s^{-1}$ of the interstellar gas that provides the effective
pressure, and $\zeta$ is the ISM metallicity, relative to the solar value. The
merger time for a remnant is
$2\times 10^6 [E_{51}^{0.32}n^{-0.47}\sigma_{10}^{-1.43}\zeta^{-0.05}]\rm yr.$
To this we add the main sequence lifetime of a $\sim 20\rm M_\odot$ star to
infer a total mixing time for the first generation of stars of about $3\times
10^6\rm yr.$  The typical volume element of interstellar gas will  be exposed,
during this initial period of marginal mixing, to at most 2-3 supernova
remnants.
The predecessor stars will not be of precisely the same mass, but will
necessarily span the range $10-100 \rm M_\odot.$ Abundance calculations for
metal-poor massive stars show that the nucleosynthetic yields are sensitive to
the progenitor mass (Woosley and Weaver 1995).
We deduce that chemical inhomogeneities will be maximal until the number
of SN remnants has increased by 1 or 2 orders of magnitude. This spans the
range in the metallicities of the oldest stars from
$\rm [Fe/H] =-4$ to $\sim -2.5,$ as noted by McWilliam $et\, al.$ (1995).

Among the many abundances determined by  McWilliam $et\, al.$ (1995),
several deserve further  consideration.
Having provided an explanation for the star-to-star dispersion
at low metallicity, we note the following abundance trends regarding two
classes of elements, the alpha elements ranging from sodium to calcium and the
iron peak elements including scandium, titanium, chromium, manganese, cobalt
and nickel.  Relative to iron, the stars with metallicity below  $\rm
[Fe/H]=-2.4$
are on average relatively rich in magnesium, calcium and cobalt, but are
relatively low in aluminium, chromium, manganese and nickel. For other
elements such as sodium, scandium and titanium, the dispersion in abundances is
large. For the alpha elements,
the trends
are consistent with what is expected, namely an overabundance of even-A nuclei
and
relative underabundance of odd-A elements because of the low  neutronization
relative to the alpha flux in very metal-poor material.
The situation is not the same  for Fe
peak nuclei, for which the relative abundances depend on the mass cut of the
supernova progenitor,
via  the temperature of the stellar material  which induces the specific
photodisintegration processes (Woosley and Weaver 1986). For most
of the stars considered, the s-process element abundances  are weak, as is
consistent with the relative lack of available neutron flux. Nevertheless,
there are conspicuous exceptions, especially for  the star CS 22898-027 with
$\rm [Fe/H]=-2.35,$ where barium and other s-process elements are overabundant
with respect to iron by factors of about 300. It is worth noting that this
specific
star has a metallicity 30 times as high as the most metal-poor stars, and
therefore there would have been  time for pre-enrichment in elements such as
$^{22}$Ne, the likely neutron source. This is
consistent with the interpretation of
this star as a CH subgiant (Thorburn and Beers 1992), for which mass
exchange may
account for enhanced s-process elemental abundances.

We
now focus attention on the three stars with the lowest ($Z/Z_\odot\approx
10^{-4}$)
abundances, analyzed by McWilliam $et\, al.$ (1995), namely CS 22885-096 ($\rm
[Fe/H]=-3.79$); CS 22949-037  ($\rm [Fe/H]=-3.99$); CD-38 245 ($\rm
[Fe/H]=-4.01$).
All three stars are relatively rich in magnesium, calcium,
scandium and cobalt relative to iron, but are relatively poor in chromium and
manganese. Nickel, aluminium and titanium track iron, as do barium and
strontium. The alpha- and heavy element abundances behave  as expected for
massive low metallicity stars. As anticipated (Table 1), iron peak elements
from scandium
to nickel show a wide range of abundances in the computations of
Woosley and Weaver (1995), depending on both the progenitor mass and on the
mass cut. These
are the only calculations we have found that study zero-metal massive star
explosions. In addition to uncertainties in the final stages of
stellar evolution,
the mass cut is questionable because the role of winds from evolved low
metallicity stars is poorly known. It seems likely that a range of masses and
mass cuts will
readily account for the abundance variations found for iron peak elements in
the extremely low metallicity stars. We must add together the yields of at
least 2, and probably no more than 3, distinct mass models,
given the average supernova remnant separation and the total mixing time for
the first generation of stars. As an illustration of this, we note that
the observed elemental abundance patterns (displayed in Table 1)
are reasonably consistent with  the variations in the mass models and mass
cuts,
for example, if we combine the yields from  13 and  22 $\rm M_\odot$ stars
(models Z13A and Z22A, respectively) with initially zero metallicity.
Model Z13A  is in agreement with the abundance pattern of Al, Cr, Mn, Co, and
Ni in CS22885-096,
whereas  the abundances of  Mg, Si, Ca, Sc and Ni in CD-38 245 are in better
accord with model Z22A. One should note that these estimates fail
to satisfactorily reproduce the observed Co/Fe overabundance.
However these observations are unexpected given that  Co is an
odd iron peak element.

Once there has been time for many generations ($e.g.$$\sim 10$) of massive
stars, we expect that these variations would
mostly be suppressed in the integrated contributions from  a range of stellar
masses. In addition, we note that the three extreme metal-poor stars show the
same spread  in strontium and barium abundances as do the more metal-rich
stars.
This means that some of the metal-poor progenitor stars should have undergone
mixing of carbon-rich,
helium-rich  and hydrogen-rich layers in order to produce N by the CN cycle,
followed by the alpha particle-induced reactions $^{14} N(\alpha,\gamma)^{18}
F(\beta^+)^{18} O(\alpha,\gamma)^{22}Ne.$

It is worth mentioning that our interpretation of the data on extremely
metal-poor stars is subject to a direct observational test.
If the stars with $\rm [Fe/H]\approx 10^{-4}$ are the outcome of the first
generation of massive stars and their associated supernovae, one would not
expect to find any stars with appreciably lower metallicity.
This is because the $\sim 10^4$ supernovae will have already produced
sufficient enrichment so that very little gas  would have retained its
primordial abundances. It is worth noting in this respect that
Norris, Peterson and Beers (1993) and  McWilliam $et\, al.$ (1995),
despite intensive
searches, find their lowest metallicity candidate stars  to be just near the
$\rm [Fe/H]=-4$ limit.
Our conclusion can be compared with that of Beers (1995), who uses
an iron  enrichment rate per supernova adopted by Primas, Molaro and
Castelli (1994) that is significantly lower than our preferred value
that we have taken  from the Woosley-Weaver (1986, 1985)
prescriptions. With the lower yield, Beers infers that
we should observe a few stars with  $\rm [Fe/H]\approx -5$ in
the next few years. While we  would advocate a higher threshold, we
can argue that in any circumstances the observed lower limit on metallicity in
the extremely metal-poor halo stars should provide a direct estimate
of the average supernova metal enrichment rate.

\bs
\goodbreak
{\bf 3. Metal-poor extragalactic systems}

Analogous reasoning can  be applied to metal-poor extragalactic systems,
including blue compact galaxies and Lyman alpha absorption line clouds seen
along the line of sight to quasars.
 Blue compact galaxies are vigorously forming stars at the present epoch, at a
rate such that
the observed level of metal enrichment implies that we may be witnessing their
first major episode of star formation and associated chemical evolution.
Consider the classic case of the most metal-poor blue compact galaxy known,
IZw18. Kunth, Lequeux, Sargent and Viallefond (1994) have reanalyzed the oxygen
and silicon abundances in the surrounding neutral gas  region. Although the
abundance determinations are uncertain, their best estimates are $O/H=7.9\times
10^{-7}$
and $Si/H=7.5\times 10^{-8}$, corresponding respectively to $10^{-3}$ and
$2\times 10^{-3}$
of the solar value. As noted by these authors, the composition of IZw18 should
be close to the primordial value, and its helium abundance should therefore
reflect the Big Bang nucleosynthesis prediction (to the extent that  one can
accurately estimate this abundance, which is still far from the case, see
$e.g.$ Pagel, Simonson, Terlevich and Edmunds 1992). The emission line region
metallicity is about
$1/30$ of the solar value and reflects the current star burst phase, which  has
presumably been underway for $\sim 10^8$ yr. Hence any ``primordial'' gas
outside the emission line region should have been contaminated only by the
first generation of stars, that we assume occurred throughout IZw18. This
would result in  an abundance level
equal to the ratio of the lifetime of the first stellar generation  ($\sim
3\times 10^6$ yr) to the age of the current star burst, which has produced the
observed emission line gas metallicity of -1.5. This results in a predicted
metallicity for the gas outside the emission line region of about
$10^{-3}$ of the solar value.

We point out that in the most metal-poor blue compact galaxy IZw18, the
relative [Si/O] ratio determined by Kunth {\it et al.}  (1994) is 0.3 (or 2 by
number). This is in agreement with the fact that alpha particle elements with
even A are produced more copiously than
O, and {\it a fortiori} Fe.  Moreover the various compilations of He versus O
and N
clearly show a large dispersion in metallicity, comparable to those observed in
low metal stars (see {\it e.g.} Pagel 1995).

A similar approach can be adopted for pregalactic clouds, such as those
observed in absorption towards high redshift quasars, where low metallicities
are measured. The most primitive clouds are those of the Lyman alpha forest,
where recent work has shown that the metallicity is about $1/300$ of the solar
value
(Cowie 1995). These are believed to be low mass clouds comparable in mass to,
and likely precursors of, dwarf galaxies, and the metallicity produced by the
first generation of stars is again expected to be on the order of
$10^{-3}$ of the solar value.

For damped Lyman alpha systems, often considered to be the precursors of disks,
comparably low metallicities are also measured in some cases, but span a wide
range.
Recently two independent teams (Songaila, Cowie, Hogan and Rugers 1994;
Carswell, Rauch, Weymann, Cooke and Webb 1994)  have reported  the
determination of a very large D/H abundance ratio, $2.4-3.0 \times 10^{-4},$ in
the direction of Q 0014+813 at $z=3.32,$  with the associated  metallicity
estimated to lie between --3.5 and --1.6.
Such a high D abundance, if confirmed, implies a  low baryon density, still
consistent with the visible matter density, at the cost, however, of producing
an excessive amount of $^3He$ (see, {\it e.g.} Palla, Galli and Silk, 1995;
Vangioni-Flam and Cass\'e 1995).
If the lower metallicity determination is correct, there has only been
enrichment by a first generation of stars,
implying a negligible level of astration.

However Tytler (1995) has reported detection of D at high redshift
in another line of sight at a level that is lower by an order of magnitude.
If astration were responsible for the low D/H value in this component, there
would have been several generations of massive stars, and the metal abundance
would be correspondingly enhanced.
It has been recently argued  that in a highly inhomogeneous early universe
dominated by primordial entropy fluctuations on scales up to those of a
galactic mass,
primordial nucleosynthesis may result in large line-of-sight variations in the
primordial D/H
ratio (Jedamzik,
Mathews  and Fuller 1995), at the price however of a high  primordial
$^7Li/H$ ratio and relatively low $^4He$ mass fraction. In this situation,
primordial conditions rather than astration are responsible for the variations
in D/H.
 Given the importance  of such an inference, further determinations of both
metallicity and deuterium abundances in  similar clouds at high redshift are of
considerable importance.


\bs
\goodbreak
{\bf 4. Conclusion}

Recent observations of extremely metal-poor halo stars
provide unique clues about the first steps of the chemical evolution history
of our galaxy. We interpret the large dispersion in elemental abundances as a
manifestation of an early phase of enrichment when there were only a few
supernova sources and mixing of their ejecta was incomplete.
A consequence of this interpretation is that we predict
the absence of any halo stars of appreciably lower metallicity.
A similar interpretation has previously been applied to account for the absence
of emission line dwarf galaxies with abundances below those of the
metal-poor compact blue galaxy IZw18 (Kunth and Sargent 1986). We have extended
this argument to account for the observed abundance differences between
Si and O (Kunth {\it et al.} 1994). Moreover the dispersion in abundances
between different
blue compact galaxies is a consequence of the earliest
phase of chemical evolution associated with the first generations of massive
stars. We have also applied
a similar argument to Lyman alpha absorption line systems at high redshift,
where we may be witnessing gas and deuterium-rich precursors to dwarf and to
disk galaxies,
corresponding to Lyman alpha forest and damped Lyman alpha systems,
respectively.

\bs

{\bf  Acknowledgements}

This research has been supported in part at the University of California,
Berkeley  by the
NSF through the Center for Particle Astrophysics,  and in part
by a grant from the D.O.E.; J. A. also wishes to gratefully acknowledge a
stimulating conversation with N. Prantzos, support as a Regents' Lecturer at
the University of California,  Berkeley, and the hospitality of the Berkeley
Department of Astronomy and the Center for Particle Astrophysics.
We also thank T. Beers and J. Truran for useful comments.
\medskip\bs
\goodbreak
\centerline {\bf References}
\smallskip
\def\pp{\parshape 3 0truecm 15truecm 0.5truecm 14.5truecm 0.5truecm 14.5
truecm}
\def\refs #1;#2;#3;#4;#5{\par\pp #1. #2, {\it #3}, {\bf #4}, #5}
\def\refb #1;#2;#3;#4{\par\pp #1. #2, {\it #3}, #4}
\parindent = 0truecm
\frenchspacing
\parskip=0.1truecm


Anders, E., and Grevesse, N. 1989, Geochim. Cosmochim. Acta, 53, 197.

Beers, T. C. 1995, in {\it The Light Element Abundances}, ed.
P. Crane (New York: Springer), 145.

Carswell, R.F., Rauch, M., Weymann, R.J., Cooke, A.J., and Webb, J.K. 1994
Mon. Not. R. astr. Soc., 268, L1.

Cioffi, D. F., McKee, C. F. and Bertschinger, E. 1988, ApJ, 334, 252.

Cowie, L. L. 1995, preprint.

Jedamzik, K., Mathews, G. J. and Fuller, G. M. 1995, ApJ, 441, 465.

Jura, M. 1986, ApJ,301, 624.

Kunth, D., Lequeux, J., Sargent, W. L. W. and Viallefond, F. 1994, Astr. Ap.,
282, 709.

Kunth, D. and Sargent, W. L. W. 1986, ApJ, 300, 496.

McWilliam, A., Preston, G., Sneden, C. and Searle, L. 1995, A. J. (in press).

Norris, J. E., Peterson, R. C. and Beers, T. C. 1993, ApJ, 415, 797.

Pagel, B. E. J. 1995, IAC Winter School on Galaxy Formation and Evolution, ed.
C. Mu\~noz-Tur\'on (Cambridge University Press, in press).

Pagel, B. E. J., Simonson, E. A., Terlevich, R. J. and Edmunds, M. G. 1992,
M.N.R.A.S., 255, 325.

Palla, F., Galli, D. and Silk, J. 1995, ApJ (in press).

 Songaila, A., Cowie, L.L., Hogan, C.J., and Rugers, M. 1994,  Nature,
368, 599.

Primas, F., Molaro, P. and Castelli, F. 1994, Astr. .Ap., 290, 885.

Thorburn and Beers, T. 1992, BAAS, 24, 1278.

Truran, J. W. and Cameron, A. G. W. 1971, Astr. Space Sci., 14, 179.

Tytler, D., 1995, ESO Meeting on Quasar Absorption lines, in press

Vangioni-Flam, E. and Cass\'e, M. 1995, ApJ, 441, 471.

Woosley, S. E. and Weaver, T. A. 1986, Ann. rev. Astr. Ap., 24, 205.

Woosley, S. E. and Weaver, T. A. 1995, ApJSuppl, in press.
\vfil\eject

\centerline{TABLE 1}
\medskip
\hrule
\medskip
\centerline{Comparison  of Abundances in 3 Extreme Metal-Poor Stars with
Supernova Models$^1$}
\medskip
\hrule
\smallskip
\hrule
\medskip
\centerline{\vbox{
\halign{ \hfil # \hfil \qquad & \hfil # \hfil \qquad & \hfil # \hfil \qquad
& \hfil # \hfil \qquad & \hfil # \hfil \qquad & \hfil # \hfil \qquad
\cr
Element  & $^{2}$CD-38 245  & $^{2}$CS 22949-037 &
$^{2}$CS 22885-096 & $^{3}$Z13A& $^{3}$Z22A  \cr
\noalign{\vskip 10pt}
 [Fe/H]&-4.01&-3.99&-3.79& -- &--\cr
\noalign{\vskip 10pt}
C\hfil      & -0.1 & 0.88 & 0.3 & -0.87 & -0.20 \cr
\noalign{\vskip 3pt}
Na\hfil     & -0.1 &  2.16 & -0.11 & -1.46 & -0.09 \cr
\noalign{\vskip 3pt}
Mg\hfil  & 0.38 & 1.20 & 0.46 & -0.77 & 0.10 \cr
\noalign{\vskip 3pt}
Al\hfil & -0.99 & -0.07 & -0.80 & -1.24 & -0.53 \cr
\noalign{\vskip 3pt}
Si\hfil & 0.66 & 0.86 & 0.69 & -0.50 & 0.24 \cr
\noalign{\vskip 3pt}
Ca\hfil & 0.58 & 0.90 & 0.58 & -0.52& 0.44 \cr
\noalign{\vskip 3pt}
Sc\hfil & 0.18 & 0.86 & 0.47 & -0.61 & 0.03 \cr
\noalign{\vskip 3pt}
Ti\hfil & 0.79 & 0.53 & 1.78 & -0.33 & -0.02\cr
\noalign{\vskip 3pt}
Cr\hfil &  -0.50 & -0.49 & -0.49 & -0.31 & 0.03 \cr
\noalign{\vskip 3pt}
Mn\hfil & -0.08 & & -0.78 & -0.67 & -0.44 \cr
\noalign{\vskip 3pt}
Co\hfil & 0.44 & 0.53 & 0.35 & 0.05 & -0.33 \cr
\noalign{\vskip 3pt}
Ni\hfil & -0.12 & & 0.83 & 0.35 & -0.10\cr
\noalign{\vskip 3pt}
Sr\hfil & -0.61 & 0.42 & -135 & & \cr
\noalign{\vskip 3pt}
Ba\hfil & 0.02 & -0.16 & & & \cr
\noalign{\vskip 8pt}
}}}
\hrule
\bigskip

$^1$ The observed stellar abundances and the model abundances, expressed with
respect to the
tabulation by Anders and Grevesse (1989),
are given logarithmically with respect to the iron abundance.

$^{2}$ Abundances from McWilliam \etal (1995).

$^{3}$  Metal-free supernova models from Woosley and Weaver (1995): models Z13A
and Z22A have   initial masses respectively of 13  and 22 M$_\odot$.
\bye